\documentclass{ws-mpla}

\def\beq{\begin{equation}}
\def\enq{\end{equation}}
\def\ba{\begin{eqnarray}}
\def\ea{\end{eqnarray}}

\def\msun{M_\odot}
\def\<{<\!\!}
\def\>{\!\!>}
\def\ra{\rightarrow}

\def\vareps{\varepsilon}

\begin{document}

\markboth{Razzaque, M\'esz\'aros, \& Waxman} {High energy
neutrinos from a slow jet model of core collapse supernovae}

%
\catchline{}{}{}{}{}
%

\title{HIGH ENERGY NEUTRINOS FROM A SLOW JET MODEL OF CORE COLLAPSE
SUPERNOVAE}

\author{\footnotesize SOEBUR RAZZAQUE}

\address{Department of Astronomy and Astrophysics, Pennsylvania State
University\\ 525 Davey Laboratory, University Park, Pennsylvania
16802, USA\\ soeb@astro.psu.edu}

\author{\footnotesize PETER M\'ESZ\'AROS}

\address{Department of Astronomy and Astrophysics, Department of Physics\\
Pennsylvania State University \\ 525 Davey Laboratory, University
Park, Pennsylvania 16802, USA}

\author{\footnotesize ELI WAXMAN}

\address{Physics Faculty, Weizman Institute of Science\\
Rehovot 76100, Israel}

\maketitle


\begin{abstract}
It has been hypothesized recently that core collapse supernovae are
triggered by mildly relativistic jets following observations of radio
properties of these explosions. Association of a jet, similar to a
gamma-ray burst jet but only slower, allows shock acceleration of
particles to high energy and non-thermal neutrino emission from a
supernova. Detection of these high energy neutrinos in upcoming
kilometer scale Cherenkov detectors may be the only direct way to
probe inside these astrophysical phenomena as electromagnetic
radiation is thermal and contains little information. Calculation of
high energy neutrino signal from a simple and slow jet model buried
inside the pre-supernova star is reviewed here. The detection prospect
of these neutrinos in water or ice detector is also discussed in this
brief review. Jetted core collapse supernovae in nearby galaxies may
provide the strongest high energy neutrino signal from point sources.

\keywords{supernovae; neutrinos; jet; core-collapse, gamma-rays: burst}
\end{abstract}


\section{Introduction}

Core collapse of massive stars which lead to supernovae (SNe) of type
Ib,c and II are in some cases associated with long duration ($\gtrsim
2$-$10^3$ s) gamma-ray bursts (GRBs), as evidenced by observed
correlations of GRB 980425/SN 1998bw, GRB 021211/SN 2002lt, GRB
030329/SN 2003dh and GRB 0131203/SN 2003lw.\cite{DellaValle:2005cr} A
relativistic jet with bulk Lorentz factor $\Gamma_{b} \gtrsim 100$,
powered by a black hole and an accretion disc which form after the
core collapse in the most likely scenario, is believed to lead to the
GRB event.\cite{MacFadyen:1998vz} Observational evidence of only a
small fraction of detected SNe associated with GRBs hints that the
frequency of highly relativistic jets in core collapse SNe is at best
1 in 1000, roughly the ratio of GRB to SN rates.\cite{Berger:2003kg}
However, a significantly larger fraction ($\lesssim 10\%$ of type Ib/c
rate\cite{Bloom:1998ad,hansen99}) of SNe (also called {\em
hypernovae}) may have mildly relativistic jets associated with
them.\cite{Macfadyen:2001,Waxman:2001kt,Ramirez-Ruiz:2002jq,Granot:2003ym}
One or more of the following observations support the jetted SN
hypothesis:
\begin{itemlist}
\item High expansion velocity (30-40 $\times 1000$ km/s) first observed in SN
1998bw.\cite{Nomoto:2001} \item Radio afterglow not associated with
$\gamma$-ray emission.\cite{Granot:2004rh} \item Asymmetric explosion
supported by polarimetry observations of SN type
Ib/c.\cite{Wang:2001,Leonard:2000}
\end{itemlist}

Numerical simulations of core collapse SNe, carried out over the last
three decades have failed to produce a successful explosion by a
prompt shock wave created due to the collapse of its iron
core.\cite{Mezzacappa:2000jj} The deposition of bulk kinetic energy in
a jet form into the stellar envelope may help disrupt and blow it up
making the SN possible.\cite{Khokhlov:1999,Macfadyen:2001} The
presence of a jet is also conducive to shock acceleration of
particles. In case of a GRB, internal shocks of plasma material along
the jet accelerate protons and electrons which radiate observed
$\gamma$-rays.\cite{Zhang:2003uk} High energy protons may escape as
cosmic rays and/or produce 100 TeV neutrinos by interacting with
$\gamma$-rays {\em in situ}.\cite{Waxman:1997ti} While the GRB jet is
making its way out of the collapsing stellar progenitor it is expected
to produce 10 TeV neutrio precursor
burst.\cite{Meszaros:2001ms,Razzaque:2003uv} These neutrinos are
emitted even in the cases when the jets do not manage to burrow
through the stellar envelope and choke inside without producing
observable $\gamma$-rays. The jets in core collapse SNe or hypernovae
which is the topic of this review are slow with $\Gamma_b \sim$ few
and choke inside the stellar envelope.\cite{Razzaque:2004yv} Neutrinos
produce from such jets are typically of a hundred GeV to TeV
energy.\cite{Razzaque:2004yv,Ando:2005xi}

As opposed to 10 MeV thermal neutrinos produced by the core collapse
SN shocks which have been detected from SN 1987A in our own
galaxy,\cite{Hirata:1987hu} \footnote{see Ref.~\refcite{Cei:2002mq}
for a review of current experimental status to detect these low energy
neutrinos.} high energy neutrinos from the jets may be detected from a
longer distance because of an increasing detection prospect with
neutrino energy. Kilometer scale ice and water Cherenkov detectors
such as IceCube\cite{Ahrens:2003ix} and ANTARES\cite{Sokalski:2005sf}
which are currently being built in Antarctica and in the Mediterranean
will have an excellent chance to detect these neutrinos from SNe
within the nearest 20 Mpc.

The organization of this brief review is as follows: In Sec.
\ref{sec:core-collapse} a basic core collapse SN picture is
outlined and a particular slow jet model in
Sec. \ref{sec:jet-model}. Shock acceleration and the maximum energy
reachable by protons are discussed in
Sec. \ref{sec:proton-acc}. Neutrino flux on Earth from a point source
and diffuse sources is calculated in Sec. \ref{sec:nu-flux} and their
detection prospects in Sec. \ref{sec:events}. Conclusions are given in
Sec. \ref{sec:summary}.

\section{Core Collapse Scenario} 
\label{sec:core-collapse}

Nuclear fusion reactions, similar to the ones which take place in our
sun, constantly enrich the interior of a star forming an iron core as
the end product. Burning up all fusion materials causes hydrodynamic
instability due to lack of radiation pressure from inside the
star. The immense gravitational pressure of the stellar envelope
and/or overlying material causes the core of stars with mass $\gtrsim
8 M_{\odot}$ to collapse at this point. The density of the compressed
core material reaches a few times the nuclear density and a rising
temperature helps iron dissociate into nucleons and alpha particles.

Infall of stellar material onto the core produces $\sim 10$ MeV
neutrinos by the process of electron capture on protons ($e^- p
\ra \nu_e n$). The density of neutrons in the core exceeds that of
protons in this process, called {\it neutronization}. Initially the
neutrinos are trapped within a radius called {\it neutrinosphere}
because of a density $\gtrsim 10^{12}$ g-cm$^{-3}$. For progenitors of
mass $\lesssim 28 M_{\odot}$, the increasing degeneracy pressure of
the neutrons leads to a rebound, which sends a shockwave through the
core. While traversing through the core, the shockwave heats up
material, dissociates more iron atoms and releases trapped $\nu_e$
from the neutrinosphere.  Neutrinos carry away $\lesssim 10^{52}$ erg
of energy or roughly $1\%$ of the total gravitational binding energy
in this bursting phase which lasts for a few milliseconds. The
shockwave, however, does not reach the envelope to drive it away
because of heavy energy loss and the star fails to explode into a
supernova.

The mechanisms envisaged to produce a successful supernova explosion,
such as observed in nature, may be divided into two main categories
despite many uncertainties such as the mass loss rate of the
pre-supernova star and neutrino transport in the core, to name a few.
The first is a {\it revived shock} model, for stars initially less
massive than $\sim 28\msun$, where the core collapses to make a
neutron star. In this case, the above-mentioned stalled supernova
shock is re-energized by neutrino absorption on nucleons outside the
stellar core (${\bar \nu}_e p \ra e^+ n$; $\nu_e n \ra e^- p$),
re-energizing them. The shock wave then reaches the envelope and
expels it away. After the supernova explosion, the stellar core cools
down in next 10's of seconds by emitting $\sim 3 \times 10^{53}$ erg
of energy in neutrinos of all flavors created by lepton pair
annihilation ($e^+e^- \ra \nu_{\mu, \tau} {\bar \nu}_{\mu, \tau}$),
neutrino pair annihilation ($\nu_e {\bar \nu}_e \ra \nu_{\mu, \tau}
{\bar \nu}_{\mu, \tau}$) and nucleon bremsstrahlung ($NN \ra NN
\nu_{\mu, \tau} {\bar \nu}_{\mu, \tau}$). These neutrinos are thermal 
with a typical energy $\sim 10$ MeV. A neutron star is left over after
the stellar core cools down, following ejection of the envelope and
outer core.

In the second scenario, which is the relevant case for the current
review, a star initially more massive than $\sim 28 \msun$ undergoes
core collapse. (a) If the initial mass is $28\msun \lesssim M_\ast
\lesssim 40\msun$, the core collapses initially to a neutron star, but
after fall-back of additional core gas which did not reach ejection
velocity, it collapses further to a black hole (BH) of mass in excess
of the Chandrasekhar mass (the neutron degeneracy pressure not being
sufficient to counteract gravity for this mass). (b) If the initial
mass $M_\ast \gtrsim 40\msun$, the core collapses directly to a black
hole of mass $\gtrsim 3-5\msun$ (see Ref. \refcite{Fryer:2003}
e.g.). Thermal neutrinos of $\sim 10-30$ MeV are also produced in the
neutronized core, before it falls into the black hole. This neutrino
luminosity, of order several solar luminosities, may or may not be
able to eject (via absorption in re-energization) the outer envelope,
which is needed in order for it to appear as supernova detectable by
its photon emission.  As in the case of stars of mass $\lesssim
28\msun$, numerical simulations have not yet been able to prove
whether stars in this mass range eject their envelope; and,
observationally, it is unclear whether any observed supernova can be
ascribed to progenitors in this mass range.  However, stars more
massive than $\sim 28\msun$ are certainly observed, and from well
understood physics, they must core-collapse to a black hole.  This is
the gist of what happens to stars of mass $\gtrsim 28\msun$ if the
stellar core is rotating slowly. A black hole, and possibly a
temporary small accretion disk is formed after core collapse, which
may not greatly affect the symmetry of the collapse and/or envelope
ejection.

The situation is thought to be drastically different for stars in the
range $M_\ast \gtrsim 28\msun$ whose core is
fast-rotating.\cite{Zhang:2002yk} In particular, for a core angular
momentum in the range (3-20)$\times 10^{16}$ cm$^2$ s$^{-1}$ for this
model, the black hole and disk accretion can serve as an energy source
for powering a long GRB, and may be able to eject a stellar envelope
with kinetic energies possibly 10-100 times higher (at least in an
isotropic-equivalent sense) than the energy of typical type Ib/c or
type II SNe. Fast core rotation also helps in forming low density
channels along the rotation axis, by centrifugal evacuation. Part of
the material accreting from the disc onto the black hole can then be
ejected as narrow jets along the axes, collimated by the gas pressure
of the envelope. The jets are powered by neutrino annihilation or
magnetohydrodynamic stresses.  The pressure and energy deposition from
the jet helps eject the star's envelope and thus the SN explosion
happens. In the case of a highly relativistic jet which manages to
break through the stellar envelope, the GRB event takes place outside
the star in an optically thin environment. A slow jet, which is
modelled in the next section, may never break through, but the ejected
material can give rise to an ``orphan'' radio afterglow which is not
associated with $\gamma$-ray emission. However, non-thermal high
energy ($\sim$ TeV) neutrinos are produced ubiquitously in both cases.

\section{A Slow Jet Model} 
\label{sec:jet-model}

\begin{figure} [t]
\centerline{\epsfxsize=3.in \epsfbox{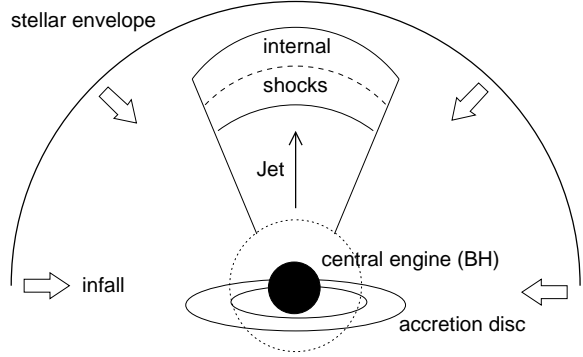}}
\caption{Cartoon of a mildly relativistic jet buried inside the 
envelope of a collapsing star. The jet may deposit its kinetic energy
and help blow up the envelope but may not get out itself.}
\label{fig:slowjet}
\end{figure}

The pre-supernova star, after losing its outermost envelope, is
typically a Wolf-Rayet star of radius $\sim 10^{11}$ cm in the case of
a type Ib or larger in the case of a type II supernova.  The mildly
relativistic SN jet may be modelled inside the pre-supernova star (see
Fig. \ref{fig:slowjet}) with a bulk Lorentz factor $\Gamma_{b} \sim
10^{0.5} \Gamma_{b,0.5}$ and a total jet kinetic energy $E_{j} \sim
10^{51.5} E_{51.5}$ ergs which is $\sim 1\%$ of the total energy
released in the SN explosion. Because of its relativistic motion, the
jet is most likely beamed with an opening half angle $\theta_{j} \sim
1/\Gamma_{b}$ which is much wider compared to a GRB jet of $\Gamma_{b}
\gtrsim 100$. Assuming a duration $t_{j} \sim 10 t_{j,1}$ s, the
isotropic equivalent kinetic luminosity of the jet is $L_{\rm kin}
\simeq 2\Gamma_{b}^{2} E_{j}/t_{j}$. With a jet variability time scale
$t_{v} \sim 0.1 t_{v,-1}$ s at the base, internal shocks between
plasma materials moving along the jet occur at a radius $r_{j} \simeq
2 \Gamma_{b}^{2} c t_{v} \approx 10^{10.8} \Gamma_{b,0.5}^2 t_{v,-1}$
cm, which is below the stellar surface.

Internal shocks convert a fraction $\vareps_{e} \sim 0.1
\vareps_{e,-1}$ of the bulk kinetic energy ($E_{j}$) into random
electron motion analogous to the GRB fireball models. In the case of
GRBs, these relativistic electrons would emit synchrotron photons in
an optically thin environment, which are observed as $\gamma$-rays on
Earth. The density of these electrons and baryons (since they are
coupled to the baryons) in the jet may be estimated as
\ba n'_{e} \simeq n'_{p} \simeq \frac{L_{\rm kin}}{4\pi r_{j}^2
\Gamma_{b}^2 m_{p}c^3} \simeq \frac{E_{j}} {2 \pi r_{j}^2 m_{p}
c^3 t_{j}} \approx 10^{20.5} \frac{E_{51.5}} {\Gamma_{b,0.5}^{4}
t_{j,1} t_{v,-1}^{2}} ~{\rm cm}^{-3}, \label{particle-density} \ea
in the comoving jet frame.\footnote{Hereafter the variables in the
comoving and in the laboratory frame or local rest frame are denoted
with and without a prime respectively.} The opacity to Thomson
scattering by photons
\ba \tau'_{\rm Th} \simeq \frac{\sigma_{\rm Th} n'_{e,p} r_{j}}
{\Gamma_{b}} \approx 10^{6.6} \frac{E_{51.5}} {\Gamma_{b,0.5}^{3}
t_{j,1} t_{v,-1}} \label{Thomson-opacity} \ea
in the comoving frame is then very high.

The magnetic fields in the jet, built up by turbulent motions in the
shock region, have an energy characterized as a fraction $\vareps_{B}
\sim 0.1 \vareps_{B,-1}$ of the total jet energy, $B^{'2}/8\pi \simeq
\vareps_{B} L_{\rm kin}/4\pi r_{j}^2 \Gamma_b^2 c$. The corresponding 
magnetic field strength in the jet is given by
\ba B' \simeq \left( \frac{4 \vareps_{B} E_{j}} {r_{j}^2 c t_{j}}
\right)^{1/2} \approx 10^{9} \left( \frac{\vareps_{B,-1} E_{51.5}}
{\Gamma_{b,0.5}^2 t_{j,1} t_{v,-1}} \right)^{1/2} ~{\rm G}.
\label{B-field} \ea
Electrons and protons are expected to be accelerated to high energies
in the internal shocks, via the Fermi mechanism. The electrons cool
down rapidly by synchrotron radiation in the presence of the magnetic
field in Eq. (\ref{B-field}). However, due to the large Thomson
optical depth in Eq. (\ref{Thomson-opacity}), these photons thermalize
and the corresponding black-body temperature is
\ba E'_{\gamma} \simeq \left( \frac{15 (\hbar c)^{3} \vareps_{e}
E_{j}} {2\pi^{4} r_{j}^2 c t_{j}} \right)^{1/4} \approx 4.3 \left(
\frac{\vareps_{e,-1} E_{51.5}}{\Gamma_{b,0.5}^4 t_{j,1}
t_{v,-1}^2} \right)^{1/4} ~{\rm keV}. \label{bb-phot-energy} \ea
The volume number density of these thermal photons may be roughly
calculated as
\ba n'_{\gamma} \simeq \frac{\vareps_{e} E_{j}}{2 \pi r_{j}^{2} c
E'_{\gamma} t_{j}} \approx 10^{24.8} \left(\frac{\vareps_{e,-1}^3
E_{51.5}^3} {\Gamma_{b,0.5}^{12} t_{j,1}^3 t_{v,-1}^6}
\right)^{1/4} ~{\rm cm}^{-3}. \label{bb-phot-density} \ea
It may be noted that photons from the shocked stellar plasma do not
diffuse into the jet due to the high optical depth $\tau'_{\rm Th}$ in
Eq. (\ref{Thomson-opacity}), and the number density in
Eq. (\ref{bb-phot-density}) is roughly constant.

\section{Proton Acceleration and Cooling Processes} 
\label{sec:proton-acc}

The shock acceleration time for a proton of energy $E'_p$ is
proportional to its Larmor's radius and may be estimated as
\ba t'_{\rm acc} \simeq \frac{A E'_{p}}{q c B'} \approx 10^{-12}
\left( \frac{E'_{p}}{\rm GeV} \right) \left( \frac{\kappa_{1}^2
\Gamma_{b,0.5}^{4} t_{j,1} t_{v,-1}^2} {\vareps_{B,-1} E_{51.5}}
\right)^{1/2} ~{\rm s}, \label{proton-acc-time} \ea
where $\kappa \sim 10\kappa_{1}$ and the magnetic field in
Eq. (\ref{B-field}) was used. The maximum proton energy is limited by
requiring this time not to exceed the dynamic time scale for the shock
to cross plasma material: $t'_{\rm dyn} \simeq t_v \Gamma_b$, or any
other possible proton cooling process time scale which we discuss
next.

\subsection{Electromagnetic cooling channels}

\subsubsection{Synchrotron and inverse Compton}

The cooling time scale for protons by synchrotron radiation in the
same magnetic field which is responsible for its acceleration is given
by
\ba t'_{\rm syn} \simeq \frac{6\pi m_{p}^{4} c^{3}} {\sigma_{\rm
Th}\beta^{2} m_{e}^{2} E'_{p} B^{'2}} \approx 3.8 \left(
\frac{E'_{p}}{\rm GeV} \right)^{-1} \left(
\frac{\Gamma_{b,0.5}^{4} t_{j,1} t_{v,-1}^{2}} {\vareps_{B,-1}
E_{51.5}} \right) ~{\rm s}, \label{proton-syn-cool} \ea
with $\beta = v/c \sim 1$. Inverse Compton (IC) scattering of thermal
electron synchrotron photons is another cooling channel for high
energy protons. The IC cooling time scale in the Thomson and
Klein-Nishina (KN) regimes, valid for $E'_p$ much less or greater than
$m_{p}^{2} c^{4}/ E'_{\gamma} \approx 10^{5.3} (\Gamma_{b,0.5}^4
t_{j,1} t_{v,-1}^2/E_{51.5} \vareps_{e,-1})^{1/4}$ GeV respectively,
as
\ba t'_{\rm IC, Th} = \frac{3 m_{p}^{4} c^{3}} {4 \sigma_{\rm Th}
m_{e}^{2} E'_{p} E'_{\gamma} n'_{\gamma}} \approx 3.8 \left(
\frac{E'_{p}}{\rm GeV} \right)^{-1} \left( \frac{
\Gamma_{b,0.5}^{4} t_{j,1} t_{v,-1}^{2}} {\vareps_{e,-1} E_{51.5}}
\right) ~{\rm s} \nonumber \\ t'_{\rm IC, KN} = \frac{3 E'_{p}
E'_{\gamma}} {4 \sigma_{\rm Th} m_{e}^{2} c^{5} n'_{\gamma}}
\approx 10^{-10.5} \left( \frac{E'_{p}}{\rm GeV} \right) \left(
\frac{\Gamma_{b,0.5}^{4} t_{j,1} t_{v,-1}^2} {\vareps_{e,-1}
E_{51.5}} \right)^{1/2} ~{\rm s}. \label{proton-IC-cool} \ea
Here we used the thermal photons with peak energy and density given in
Eqs. (\ref{bb-phot-energy}) and (\ref{bb-phot-density}) respectively
as targets.

\subsubsection{Bethe-Heitler}

Because of a high density of thermal photons in the SN jet in
Eq. (\ref{bb-phot-density}), protons may produce $e^+e^-$ pairs by
interacting with them, a process known as Bethe-Heitler (BH). The
cross-section for BH interaction: $p\gamma \ra p e^+ e^-$ is given by
$\sigma_{\rm BH} = \alpha r_{e}^{2} \{(28/9) {\rm ln}[2E'_{p}
E'_{\gamma}/ (m_{p} m_{e} c^{4})]-106/9 \}$. The logarithmic increase
of the cross-section with incident proton energy implies that this is
a very efficient cooling mechanism for high energy protons. The $e^+
e^-$ pairs are produced at rest in the center of mass (c.m.) frame of
the collision and acquire an energy $m_{e} c^{2} \gamma'_{\rm c.m.}$
each in the comoving frame. Here $\gamma'_{\rm c.m.} = (E'_{p} +
E'_{\gamma})/ (m_{p}^{2} c^{4} + 2 E'_{p} E'_{\gamma})^{1/2}$ is the
Lorentz boost factor of the c.m. in the comoving frame. The energy
lost by the proton in each BH interaction is thus the energy of the
created pairs in the comoving frame, and is given by $\Delta E'_{p} =
2 m_{e} c^{2} \gamma'_{\rm c.m.}$. The energy loss rate of the proton
is proportional to the BH scattering rate as given by $dE'_p/dt'_{\rm
BH} = n'_{\gamma} c \sigma_{\rm BH} \Delta E'_{p}$, and the
corresponding proton cooling time is
\ba t'_{\rm BH} = \frac{E'_{p}} {dE'_{p}/dt'_{\rm BH}} =
\frac{E'_{p} (m_{p}^{2} c^{4} + 2 E'_{p} E'_{\gamma})^{1/2}} {2
n'_{\gamma} \sigma_{\rm BH} m_{e} c^{3} (E'_{p} + E'_{\gamma})},
\label{proton-BH-cool}\ea
in the comoving frame.

\subsection{Hadronic cooling channels}

Photomeson ($p\gamma$) and proton-proton ($pp$) interactions which are
responsible for producing high energy neutrinos may also serve as a
cooling mechanism for the shock accelerated protons. The $p\gamma$ at
the $\Delta^+$ resonance and the average $pp$ cross-sections are
$\sigma_{p\gamma} = 5\times 10^{-28}$ cm$^2$ and $\sigma_{pp} \approx
5\times 10^{-26}$ cm$^2$ respectively.  The corresponding optical
depths, given by
\ba \tau'_{p\gamma} &=& \frac{\sigma_{p\gamma} n'_{\gamma}
r_j}{\Gamma_b} \approx 10^{7.8} \left( \frac{\vareps_{e,-1}^3
E_{51.5}^3} {\Gamma_{b,0.5}^8 t_{j,1}^3 t_{v,-1}^2} \right)^{1/4}
~{\rm and} \nonumber \\ \tau'_{pp} &=& \frac{\sigma_{pp} n'_{p}
r_j}{\Gamma_b} \approx 10^{5.5} \left( \frac{E_{51.5}}
{\Gamma_{b,0.5}^3 t_{j,1} t_{v,-1}} \right), \label{p-opacities} \ea
are very high. The threshold proton energy for $\Delta^+$ production
against the target thermal photons of energy $E'_{\gamma}$ in
Eq. (\ref{bb-phot-energy}) is
\ba E'_{p, \Delta^+} &=& \frac{0.3 {\rm GeV}^2}{E'_{\gamma}}
\approx 10^{4.8} \left(  \frac{\Gamma_{b,0.5}^4 t_{j,1}
t_{v,-1}^2} {\vareps_{e,-1} E_{51.5}} \right)^{1/4} ~{\rm GeV.}
\label{pg-energy} \ea

Adopting the energy loss by a proton $\Delta E'_p \approx 0.2 E'_p$
and $0.8 E'_p$ respectively for each $p\gamma$ and $pp$ interaction,
the hadronic cooling time scales are
\ba t'_{p\gamma} &=& \frac{E'_p}{c \sigma_{p\gamma} n'_{\gamma}
\Delta E'_p} \approx 10^{-7.3} \left(\frac{\Gamma_{b,0.5}^{12}
t_{j,1}^3 t_{v,-1}^6} {\vareps_{e,-1}^3 E_{51.5}^3} \right)^{1/4}
~{\rm s} \nonumber \\ t'_{pp} &=& \frac{E'_p}{c \sigma_{pp} n'_{p}
\Delta E'_p} \approx 10^{-5.6} \left( \frac{\Gamma_{b,0.5}^{4} t_{j,1}
t_{v,-1}^{2}} {E_{51.5}} \right) ~{\rm s,} 
\label{proton-had-cool} \ea
using Eqs. (\ref{bb-phot-density}) and (\ref{particle-density}). The
photomeson cooling time scale $t'_{p\gamma}$ above is roughly valid at
$E'_p = E'_{p, \Delta^+}$ and one needs to use the thermal photon
spectrum to calculate it for different proton energies.

\subsection{Maximum proton energy}

\begin{figure} [t]
\centerline{\epsfxsize=3.75in \epsfbox{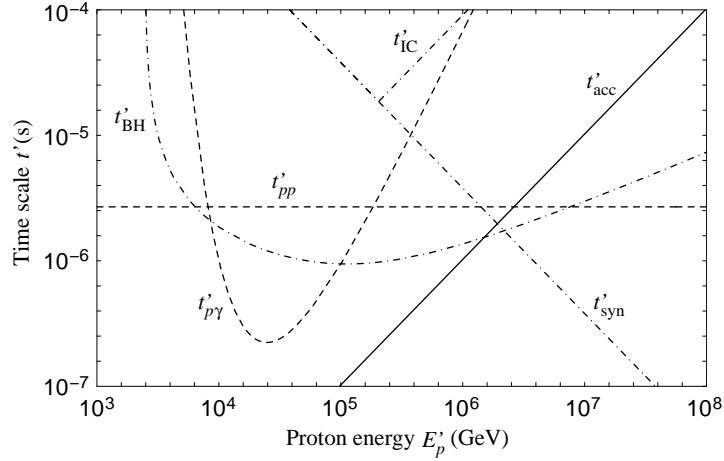}}
\caption{Proton cooling time scales for different hadronic 
(dashed lines) and electromagnetic (dot-dashed lines) processes:
photomeson ($t'_{p\gamma}$), proton-proton ($t'_{pp}$), synchrotron
radiation ($t'_{\rm syn}$), IC scattering ($t'_{\rm IC}$) and BH
($t'_{\rm BH}$); in the comoving frame, as functions of energy. The
shock acceleration time ($t'_{\rm acc}$) is plotted with a solid
line. The parameters used to model the SN jet are $E_{\rm
sn}=10^{51.5}$ erg, $t_{j} = 10$ s, $t_{v} = 10^{-1}$ s, $\Gamma_{b} =
10^{0.5}$ and $\vareps_{e} = \vareps_{B} = 0.1$.}
\label{fig:proton-cool-time}
\end{figure}

The shock acceleration time in Eq. (\ref{proton-acc-time}) for protons
in the SN jet (solid line) and the different cooling time scales are
plotted in Fig. \ref{fig:proton-cool-time} as functions of the
comoving proton energy. The dashed lines are hadronic cooling time
scales in Eq. (\ref{proton-had-cool})\footnote{A thermal photon
spectrum is used to calculate $p\gamma$ cooling with a delta-function
approximation.} and the dot-dashed lines are electromagnetic cooling
time scales in Eqs. (\ref{proton-syn-cool}), (\ref{proton-IC-cool})
and (\ref{proton-BH-cool}). Note that the hadronic cooling time
($t'_{pp}$), the BH cooling time ($t'_{\rm BH}$) and the synchrotron
cooling time ($t'_{\rm syn}$) are first longer and then shorter than
the maximum proton acceleration time ($t'_{\rm acc}$). The hadronic
($t'_{p\gamma}$) and IC scattering ($t'_{\rm IC}$) are not efficient
cooling mechanisms for protons. The maximum proton energy can be
roughly estimated, by equating the $t'_{\rm syn}$ to $t'_{\rm acc}$,
from Eqs. (\ref{proton-acc-time}) and (\ref{proton-syn-cool}) as
\ba E'_{p,{\rm max}} = \left( \frac{6\pi m_p^4 c^4 q \Gamma_b^2}
{\sigma_{\rm Th} m_e^2 B'} \right)^{1/2} \approx 10^{6.3} \left(
\frac{\Gamma_{b,0.5}^4 t_{j,1} t_{v,-1}^2}{\kappa_{1} \vareps_{B,-1}
E_{51.5}} \right)^{1/4} ~{\rm GeV}, \label{max-proton-energy} \ea
since $t'_{\rm BH} \approx t'_{\rm syn} \approx t'_{\rm acc}$ at this
energy.

\section{Neutrino Production and Flux on Earth} 
\label{sec:nu-flux}

Shock accelerated protons in the SN jet can produce non-thermal
neutrinos by photomeson ($p\gamma$) interactions with thermal
synchrotron photons and/or by proton-proton ($pp$) interactions with
cold protons present in the shock region. As shown in
Fig. \ref{fig:proton-cool-time}, the $p\gamma$ process is dominant in
the energy range $E'_p \approx 10^4$ - $10^{5.2}$ GeV and the $pp$
process is dominant at other energies.

In the case of $p\gamma$ interactions at the $\Delta^+$ resonance,
neutrinos are produced from charged pion ($\pi^+$) decay as $p \gamma
\ra \Delta^+ \ra n \pi^{+} \ra \mu^{+} \nu_{\mu} \ra e^{+} \nu_e {\bar
\nu}_{\mu} \nu_{\mu}$. The $pp$ interactions also produce charged 
pions ($\pi^{\pm}$) and kaons ($K^{\pm}$) and their decay modes are
the same as above with $100\%$ and $63\%$ branching ratios
respectively.\footnote{$\pi^-$ and $K^-$ decay modes are charge
conjugate of $\pi^+$ and $K^+$ decay modes.} The total pion (kaon)
multiplicity in each $pp$ interaction is $\sim 1$ ($\sim 0.1$) in the
energy range considered here.\cite{Razzaque:2003uw} The energy of the
shock accelerated protons in the SN jet is expected to be distributed
as $\propto 1/E_p^{'2}$, following the standard shock acceleration
models. Charged mesons, produced by $pp$ and $p\gamma$ interactions,
are expected to follow the proton spectrum with $\sim 20\%$ of the
proton energy for each pion or kaon.

\subsection{Meson suppression factors}

High-energy pions, kaons and muons produced by $p\gamma$ and $pp$
interactions do not all decay to neutrinos as electromagnetic
(synchrotron radiation and IC scattering) and hadronic ($\pi p$ and
$Kp$ interactions) cooling mechanisms reduce their energy. Muons are
severely suppressed by electromagnetic energy losses and do not
contribute much to high-energy neutrino production. Suppression
factors for pion and kaon decay neutrinos are discussed next.

The synchrotron and IC cooling times may be combined into a single
electromagnetic cooling rate as $t_{\rm em}^{'-1} = t_{\rm syn}^{'-1}
+ t_{\rm IC}^{'-1}$. For IC cooling in the Thomson regime $t'_{\rm IC}
\approx t'_{\rm syn}$ and in the KN regime $t'_{\rm IC} \gg t'_{\rm
syn}$. The electromagnetic cooling time scales for mesons may be
estimated assuming $t'_{\rm em} \approx t'_{\rm syn}$ for simplicity
as
\ba t'_{\rm em} \approx \cases{ 0.002 ~(E'_{\pi}/{\rm GeV})^{-1} 
\cr 0.3 ~(E'_{K}/{\rm GeV})^{-1} } \times \left(
\frac{\Gamma_{b,0.5}^{4} t_{j,1} t_{v,-1}^{2}} {\vareps_{B,-1}
E_{51.5}} \right) ~{\rm s.} \label{meson-emcool-time} \ea
The hadronic energy losses for mesons is similar to the proton energy
losses by $pp$ interactions in Eq. (\ref{proton-had-cool}) with the
same $\pi p$ and $Kp$ cross-section of $\approx 3\times 10^{-26}$
cm$^2$. The corresponding hadronic cooling time scales for mesons are
\ba t'_{\rm had} \approx 10^{-5.4} \left( \frac{\Gamma_{b,0.5}^{4} 
t_{j,1} t_{v,-1}^{2}} {E_{51.5}} \right) ~{\rm s,}
\label{meson-hadcool-time} \ea
with $\Delta E' = 0.8 E'.$

\begin{figure} [t]
\centerline{\epsfxsize=3.75in \epsfbox{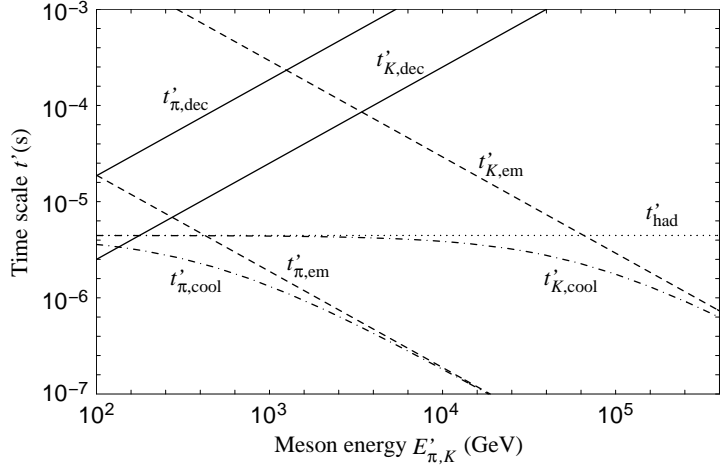}}
\caption{The Hadronic ($t'_{\rm had}$), electromagnetic 
($t'_{\pi;K,\rm em}$) and total ($t'_{\pi;K,\rm cool}$) cooling time
scales for mesons: $\pi$ and $K$ as functions of comoving energy in a
SN jet model by dashed, dotted and dot-dashed lines respectively. The
meson decay time scales boosted by the respective Lorentz factors
($t'_{\pi;K,\rm dec}$) are also plotted with solid lines. The
parameters used to model the SN jet are $E_{\rm sn}=10^{51.5}$ erg,
$t_{j} = 10$ s, $t_{v} = 10^{-1}$ s, $\Gamma_{b} = 10^{0.5}$ and
$\vareps_{e} = \vareps_{B} = 0.1$.}
\label{fig:meson-cool-time}
\end{figure}

The electromagnetic ($t'_{\pi;K,\rm em}$) and hadronic ($t'_{\rm
had}$) cooling time scales for mesons along with their decay times
boosted by the respective Lorentz factors, $t'_{\pi;K,\rm dec}$, are
plotted in Fig. \ref{fig:meson-cool-time}. The total cooling time
scale $t'_{\pi;K,\rm cool} = 1/(t_{\pi;K,\rm em}^{'-1} + t_{\pi;K,\rm
had}^{'-1})$ is first dominated by the hadronic and then by the
electromagnetic cooling channel. The ratio $t'_{\pi;K,\rm
cool}/t'_{\pi;K,\rm dec}$ determines the suppression of mesons before
they decay to neutrinos.

From the condition $t'_{\rm em} = t'_{\rm had}$, one may roughly
define a break energy as
\ba E'_{\rm br} = \cases{ 423/\vareps_{B,-1} ~{\rm GeV} & (pion) \cr
6.5\times 10^4/\vareps_{B,-1} ~{\rm GeV} & (kaon),}
\label{meson-break-energy} \ea
above and below which the mesons cool by electromagnetic and hadronic
interactions respectively. The corresponding suppression factor may be
defined from the ratios $t'_{\rm em}/t'_{\rm dec}$ and $t'_{\rm
had}/t'_{\rm dec}$ as
\ba \zeta = \cases{ 10^{-1.25} (E'_{\pi}/E'_{\pi,\rm br})^{-\beta} 
\vareps_{B,-1} \cr 10^{-2.57} (E'_{K}/E'_{K,\rm br})^{-\beta} 
\vareps_{B,-1} } ~;~ \beta = \cases{1 ~;~ E'_{\pi,K} < E'_{\pi;K,\rm br} 
\cr 2 ~;~ E'_{\pi,K} > E'_{\pi;K,\rm br}. } 
\label{suppression-factor} \ea
Another break energy may be defined from the condition $t'_{\rm dec} =
t'_{\rm had}$ below which mesons decay to neutrinos without any
suppression. However, this energy: $24$ and $177$ GeV respectively for
pions and kaons in the comoving frame is low. The observed energy on
Earth in both cases is below the detection threshold energy as will be
discussed shortly.

\subsection{Neutrino fluence from individual sources}

If the shock accelerated protons could travel unimpeded from the SN
jet at a luminosity distance $D_L$, then their isotropic equivalent
fluence on Earth would be\footnote{Denoting the quantities in the
observer's frame with subscript ``ob''.}
\ba {\cal F}_{p,{\rm ob}} = \frac{E_{j} \Gamma_{b}^{2} (1+z)}
{2\pi D_{L}^{2} E_{p,{\rm ob}}^{2} {\rm ln}[E_{p, {\rm ob,max}}/E_{p,
{\rm ob,min}}]}. \label{proton-flux} \ea
Here $E_{p,{\rm ob}} = E_{p}/(1+z)$ and $t_{j,{\rm ob}} = t_{j}(1+z)$
are the energy and time related in the observer's and local rest
frames for the source location at redshift $z$. Of course, defelction
by magnetic fields as well as other interactions along the way prevent
such direct arrivale of the protons.

Nearly all protons are expected to convert into $\pi^+$ ($\pi^{\pm}$)
by $p\gamma$ ($pp$) interactions in the SN jet because
$\tau'_{p\gamma} \gg 1$ ($\tau'_{pp} \gg 1$) in
Eq. (\ref{p-opacities}) and assuming the charged pion multiplicity is
$\sim 1$ from $pp$ interactions as a conservative estimate. With a
$100\%$ branching ratio for pion decay to neutrinos, one may define a
multiplicative factor $f_{\pi} \sim 1$ for protons which will produce
neutrinos via pions. A similar factor $f_{K} \sim 0.063$ may be
defined for kaons as a product of $K^{\pm}$ multiplicity of $\sim 0.1$
from $pp$ interactions and $63\%$ branching ratio for kaon decay to
neutrinos. Protons produce neutral pions (kaons) and charged pions
(kaons) with equal probability in both the $p\gamma$ and $pp$
interactions.\footnote{Kaons are produced by $pp$ interactions only.}
For simplicity one may also assume that a $\nu_{\mu}$ carries 1/4 of
the charged pion (kaon) energy: $E_{\nu} \approx E_{\pi,K}/4$, from
roughly equipartition of energy between the final decay products of
$\pi^{\pm}$ and $K^{\pm}$. From these considerations one may estimate
the neutrino (of one flavor) fluence on Earth as
\ba {\cal F}_{\nu,{\rm ob}} &=& \frac{1}{8} \frac{ f_{\pi,K} 
\eta _{\pi,K} E_{j} \Gamma_{b}^{2} (1+z)} 
{2\pi D_{L}^{2} E_{\nu,\rm ob}^{2} {\rm ln}
\left[E_{\nu,\rm ob,max}/E_{\nu,\rm ob,min} \right]} 
\left( \frac{E_{\nu,\rm ob}[1+z]} 
{E_{\nu,\pi;K, \rm br}} \right)^{-\beta},
\label{nu-fluence-formula} \ea
per SN burst using Eqs. (\ref{suppression-factor}) and
(\ref{proton-flux}). Here $\beta =2$ ($1$) for $E_{\nu,\rm ob}(1+z)$
greater (less) than $E_{\nu,\pi;K,\rm br} = \Gamma_b E'_{\pi;K,\rm
br}/4$ from Eq. (\ref{meson-break-energy}). The pre-factors in
Eq. (\ref{suppression-factor}) are represented by the parameters
$\eta_{\pi} = 10^{-1.25}$ and $\eta_K = 10^{-2.57}$ respectively for
pions and kaons. For a typical ice Cherenkov detector such as IceCube,
the threshold neutrino detection energy is $E_{\nu,\rm th} \sim
10^{2.5} E_{\nu,2.5}$ GeV. The neutrino energy range is then $E_{\nu,
{\rm ob, min}}$ - $E_{\nu, {\rm ob, max}} = E_{\nu, {\rm th}}$ - $0.05
\Gamma_{b} E'_{p, {\rm max}}/(1+z) \approx 10^{2.5}$ -
$10^{5.5}/(1+z)$ GeV in the observer's frame on Earth.

As mentioned earlier, $\pi^+$'s are produced by $pp$ or $p\gamma$
interactions by shock accelerated protons of all energies. In this
case $\pi^+$ decay $\nu_{\mu}$ fluence from a SN at a distance $\sim
20$ Mpc ($D_{L} \approx 10^{25.8}$ cm, $z\sim 0$), e.g. from the Virgo
cluster, would be
\ba {\cal F}_{\nu,\pi,{\rm ob}}^{*} &\approx & 10^{-5} 
\left(\frac{E_{\nu,{\rm ob}}}{10^{2.5}~{\rm GeV}} \right)^{-4} 
~ \frac{ E_{51.5} \Gamma_{b,0.5}^2} {D_{25.8}^{2} \vareps_{B,-1}} ~
{\rm GeV^{-1}cm^{-2}}; \nonumber \\ &&
~~~~~~~~~~~~~~~~~~~~~~~~~~~10^{2.5}
\lesssim E_{\nu,{\rm ob}}/{\rm GeV} \lesssim 10^{5.5},
\label{nuflux-pi-numbers} \ea
from Eq. (\ref{nu-fluence-formula}) with $f_{\pi} =1$. The neutrino
break energy from pion decay: $E_{\nu,\pi,\rm br} \approx E_{\nu,\rm
th} \approx 10^{2.5}$ GeV in this case. A similar expression may be
derived for neutrino fluence from kaon decays. However, $pp$
interactions are overwhelmed by $p\gamma$ interactions in the energy
range $E'_p \approx 10^4$-$10^{5.2}$ GeV
(Fig. \ref{fig:proton-cool-time}). Hence kaon and the corresponding
decay neutrino production is expected to be suppressed in the energy
range $E_{\nu} \approx 10^{3.9}$-$10^{5.1}$ GeV. The neutrino break
energy from Eq. (\ref{meson-break-energy}) is $E_{\nu,K,\rm br}
\approx 10^{4.7}$ GeV and the fluence is
\ba {\cal F}_{\nu,K,{\rm ob}}^{*} &\approx & 10^{-11.7} 
\left(\frac{E_{\nu,{\rm ob}}}{10^{4.7} 
~{\rm GeV}} \right)^{-(\beta+2)} ~ \frac{ E_{51.5} \Gamma_{b,0.5}^2}
{D_{25.8}^{2} \vareps_{B,-1}^{\beta-1} } ~ {\rm GeV^{-1}cm^{-2}};
\nonumber \\ && ~~~~~~~~~~~~~~~~~~~~~~\beta = \cases{1 ~;~ 10^{2.5} 
\lesssim E_{\nu,{\rm ob}}/{\rm GeV} \lesssim 10^{3.9} \cr 
2 ~;~ 10^{5.1} \lesssim E_{\nu,{\rm ob}}/{\rm GeV}
\lesssim 10^{5.5}, }
\label{nuflux-K-numbers} \ea
on Earth from kaon decays in a SN jet at a distance $\sim 20$ Mpc.

\subsection{Diffuse neutrino flux}

The diffuse neutrino flux is calculated by summing over fluences from
all slow-jet endowed SNe distributed over cosmological distances in
Hubble time.  The SNe rate follows closely the star formation rate
(SFR) which can be modeled, as a function of redshift per unit
comoving volume,\cite{Porciani:2000ag} as
\ba 
{\dot \rho}_{*} (z) = \frac{0.32 h_{70} ~{\rm exp}(3.4 z)} {{\rm
exp}(3.8 z) + 45} ~M_{\odot} ~{\rm yr}^{-1} ~{\rm Mpc}^{-3}.
\label{SFR} 
\ea
Here $H_{0} = 70 h_{70}$ km s$^{-1}$ Mpc$^{-1}$ is the Hubble
constant.  For a Friedmann-Robertson-Walker universe, the comoving
volume element is
\ba
\frac{dV}{dz} = \frac{4\pi D_{L}^2 c}{1+z}
\left| \frac{dt}{dz} \right|,
\label{volume-element}
\ea
and the relation between $z$ and the cosmic time $t$ is
\ba
(dt/dz)^{-1} = -H_{0} (1+z) \sqrt{ (1+\Omega_{m}z) (1+z)^{2} -
\Omega_{\Lambda} (2z+z^{2})}.
\label{redshift-time}
\ea
For the standard $\Lambda$CDM cosmology, $\Omega_{m} = 0.3$ and
$\Omega_{\Lambda} = 0.7$.

The number of SNe per unit star forming mass ($f_{\rm sn}$) depends on
the initial mass function and the threshold for stellar mass to
produce SN ($M_\ast \sim 8$ M$_{\odot}$). A Salpeter model $\phi (M)
\propto M^{-\alpha}$ with different power-law indices can generate
different values for $f_{\rm sn}$, e.g. $\approx 0.0122$
M$_{\odot}^{-1}$ for $\alpha = 2.35$ and $\approx 8\times 10^{-3}$
M$_{\odot}^{-1}$ for $\alpha =
1.35$.\cite{Porciani:2000ag,Hernquist:2002rg} We adopt the model in
Ref. \refcite{Porciani:2000ag} which corresponds to the local type II
SNe rate ${\dot n}_{\rm sn}(z=0) = f_{\rm sn} {\dot \rho}_{*} (z=0)
\approx 1.2 \times 10^{-4} h_{70}^{3}$ yr$^{-1}$ Mpc$^{-3}$ agreeing 
with data.\cite{Madau:1998dg}

The distribution of SNe per unit cosmic time $t$ and solid angle
$\Omega$ covered on the sky can be written as
\ba
\frac{d^2 N_{\rm sn}}{dt d\Omega} = \frac{ {\dot n}_{\rm sn} (z)}
{4\pi} \frac{dV}{dz} = \frac{ {\dot n}_{\rm sn}(z) D_{L}^{2} c}
{(1+z)^{2}} \left| \frac{dt}{dz} \right|.
\label{sn-dist} \ea
We assume a fraction $\xi_{\rm sn} \lesssim 1$ of all SNe involve jets
and a fraction $1/2\Gamma_{b}^2$ of all such jets are pointing towards
us. The observed diffuse SNe neutrino flux, using Eqs.
(\ref{nu-fluence-formula}) and (\ref{sn-dist}), is then
\ba
\Phi_{\nu, {\rm ob}}^{\rm diff} &=& \frac{\xi_{\rm
sn}}{2\Gamma_{b}^2} \int_{0}^{\infty} dz ~ \frac{d^2 N_{\rm sn}}{dt
d\Omega} ~{\cal F}_{\nu,{\rm ob}} (E_{\nu,{\rm ob}}) \nonumber \\ &=&
\frac{\xi_{\rm sn} f_{\pi,K} \eta _{\pi,K} }{32\pi}
\frac{c E_{j}} {E_{\nu,{\rm ob}}^{2}} \int_{0}^{\infty}
dz ~\frac{{\dot n}_{\rm sn}(z)/(1+z)} {{\rm ln} \left[ E_{\nu,
\rm{ob,max}}/E_{\nu, {\rm th}} \right]}
~\left| \frac{dt}{dz} \right| \left( \frac{E_{\nu, {\rm
ob}}[1+z]}{E_{\nu,\pi;K,\rm br}} \right)^{-\beta} \nonumber \\ &&
~~~~~~~~~~~~~~~~~~~~~~~
\times ~\Theta \left( E_{\nu, {\rm th}} \le E_{\nu,{\rm ob}} \le
E_{\nu, \rm{ob,max}} \right).
\label{diff-nu-flux} \ea
with $\beta$ specified as before.

\begin{figure} [t]
\centerline{\epsfxsize=3.75in \epsfbox{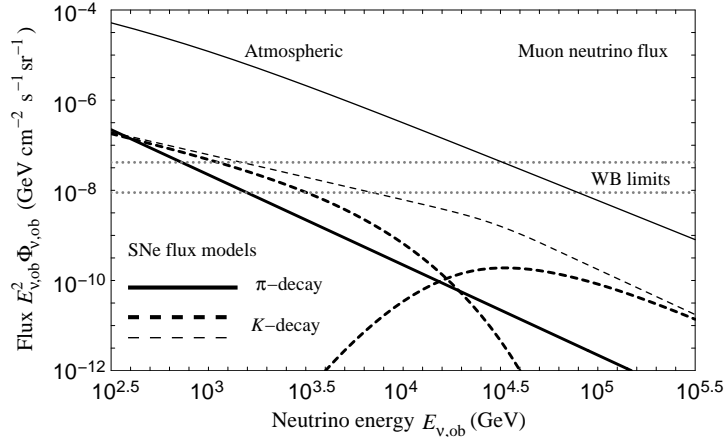}}
\caption{Diffuse muon neutrino ($\nu_{\mu}$) flux models from core 
collapse SNe assuming all of them [$\xi_{\rm sn}=1$ in
Eq. (\ref{diff-nu-flux})] are endowed with slow jets as discussed
here. The pion decay neutrinos are produced via $pp$ and $p\gamma$
interactions over the whole range of shock accelerated proton
energy. The kaon decay neutrinos are produced via $pp$ interactions
only and may be suppressed (thick dashed line) where $p\gamma$
interactions are dominant over $pp$ interactions. The thin dashed line
corresponds to the kaon decay neutrino flux if $pp$ is dominant over
the whole proton energy range. The parameters used to model the SN jet
are the same as in Figs. \ref{fig:proton-cool-time} and
\ref{fig:meson-cool-time}. Diffuse fluxes are unlikely to be detected 
because of the huge atmospheric background.}
\label{fig:nuflux}
\end{figure}

The diffuse $\nu_{\mu}$ flux from all cosmological slow-jet SNe is
plotted in Fig. \ref{fig:nuflux} by numerically integrating Eq.
(\ref{diff-nu-flux}), assuming the maximal fraction $\xi_{\rm sn}=1$.
The pion (kaon) decay flux is plotted with thick solid (dashed)
line(s). The exponential suppression in the kaon decay $\nu_{\mu}$
flux curves is due to the lack of kaon production by $pp$
interactions, as $p\gamma$ interactions, which do not produce kaons,
are dominant in the particular proton energy range corresponding to
this neutrino energy range, $E_{\nu,\rm ob} (1+z) \approx
10^{3.9}$-$10^{5.1}$ GeV. The light dashed curve corresponds to the
kaon decay $\nu_{\mu}$ flux if $pp$ interaction dominates over the
whole proton energy range for comparison. The kaon decay neutrino
fluxes are first smaller and then larger than the pion decay neutrino
flux. This is because of the kaon's heavier mass and shorter decay
time compared to pions (see Fig. \ref{fig:meson-cool-time}).

The atmospheric $\nu_{\mu}$, ${\bar \nu}_{\mu}$ flux from pion and
kaon decays (conventional flux) compatible with AMANDA
data\cite{Ahrens:2002gq} is plotted with the following
parametrization\cite{Thunman:1995}
\ba \Phi_{\nu, {\rm ob}}^{\rm atm} = \cases{ 0.012 E_{\nu,{\rm
ob}}^{-2.74}/ (1+0.002 E_{\nu,{\rm ob}}) ~;~ E_{\nu,{\rm ob}} <
10^{5.8}~{\rm GeV} \cr 3.8 E_{\nu,{\rm ob}}^{-3.17}/ (1+0.002
E_{\nu,{\rm ob}}) ~;~ E_{\nu,{\rm ob}} > 10^{5.8}~{\rm GeV}. }
\label{atmo-nu-flux} \ea
Also shown are the cosmic ray bounds (WB limits) on the diffuse
neutrino flux.\cite{Waxman:1997ti} It is unlikely that neutrino
detectors can measure the SNe diffuse fluxes plotted in Figure
\ref{fig:nuflux}, as they are below the atmospheric background. 
However, individual SNe in nearby galaxies may be detectable, 
as discussed in the next section.

\section{Neutrino Detection Prospects} \label{sec:events}

There are approximately 4000 galaxies known within 20 Mpc distance. At
the standard rate of 1 SNu = $10^{-2}$/yr/$10^{10}$ blue solar
luminosity for average galaxies, the estimated SN rate is $\gtrsim
1$/yr. The SN rate in the starburst galaxies, such as M82 and NGC253
(at distances 3.2 and 2.5 Mpc in the Northern and Southern sky,
respectively) is $\sim 0.1$/yr, much larger than in the Milkyway or in
the Magellanic clouds.\cite{Hherrero:2003} Very strong neutrino
signals in future kilometer scale neutrino detectors are expected from
these nearby SNe, over a negligible atmospheric background, using
temporal and positional coincidences with optical detections.

The directional sensitivity of the Cherenkov detectors is best for
muon neutrinos, which create muons by charge current neutrino-nucleon
($\nu N \ra \mu X$) interactions. Muons carry $\sim 80\%$ of the
incident neutrino energy. It emits Cherenkov light as it travels
faster than the speed of light in the detection media. Photo
multiplier tubes (PMTs) buried in the medium (ice, e.g. in the case of
IceCube) can detect muons by gathering their Cherenkov light. The
effective detection area of a Cherenkov detector depends on the
arrival direction of the neutrino and the energy of the muon. In
principle it can be larger than the geometrical area ($A_{\rm det}
\sim {\rm km}^2$ in the case of IceCube e.g.) as muons can be produced
outside the instrumented volume and travel inside. To achieve a good
pointing resolution a muon should {\it hit} at least 4 PMTs, for
reconstructing its track unamiguously, strung on different vertical
{\it strings}. The PMT spacing is 17 m vertically and 125 m
horizontally. With a muon energy loss of $\sim 0.2$ GeV/m and a PMT
efficiency of $\sim 30\%$, the threshold energy for muon detection is
$(3\times 125~{\rm m})\times (0.2~{\rm GeV/m})/0.3 \approx 250$ GeV as
a conservative estimate. This corresponds to a neutrino threshold
energy $E_{\nu,\rm th} \approx 10^{2.5}$ GeV as used before. The
IceCube detector is expected to have an angular sensitivity of
$1^{\circ}$ for muon tracks of energy $\lesssim 1$ TeV coming from a
zenith angle $<140^{\circ}$ and it gets better at higher
energy.\cite{Ahrens:2003ix}

\begin{figure} [t]
\centerline{\epsfxsize=3.75in \epsfbox{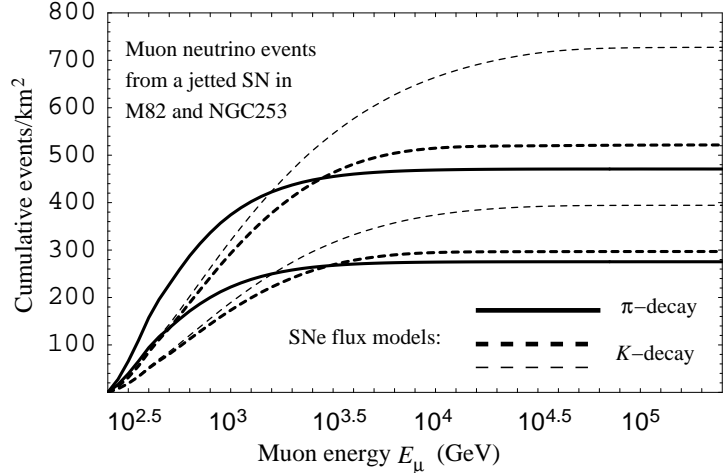}}
\caption{Cumulative muon events from $\nu_{\mu}$ flux models from 
a jetted core collapse SN in M82 and NGC253 (lower and upper sets of
lines) as functions of observed muon energy in IceCube. The pion and
kaon decay neutrino flux models are described in
Eqs. (\ref{nuflux-pi-numbers}), (\ref{nuflux-K-numbers}) and in the
caption of Fig. \ref{fig:nuflux} (also in the main text). The neutrino
events from pion decay flux models are larger (smaller) than the kaon
decay flux models at lower (higher) energy. This is because more pions
are produced than kaons but pion decay to neutrino is suppressed due
to pion energy loss at higher energy compared to kaons with heavier
mass and shorter decay time (Fig. \ref{fig:meson-cool-time}).}
\label{fig:nuevents}
\end{figure}

The number of muon events from a nearby jetted SN may be calculated as
\ba
N_{\mu} = A_{\rm det} \int_{E_{\nu,\rm th}}^{E_{\nu,\rm max}} {\cal P}
(E_{\nu}, \theta) {\cal F}_{\nu,\pi;K}^{*} (E_{\nu}) dE_{\nu},
\label{event-rate} \ea
where the neutrino fluences are given in
Eqs. (\ref{nuflux-pi-numbers}) and (\ref{nuflux-K-numbers}). The full
detection probability ${\cal P} (E_{\nu}, \theta)$ depends on the
source's angular position ($\theta$) and hence on the Earth's
shadowing effect, as well as the energy dependent $\nu N$ cross
section.\cite{Razzaque:2003uw} The cumulative number of muon events
from M82 and NGC253 are plotted in Fig. \ref{fig:nuevents} per
individual SN with its jet pointing towards Earth . The $\nu_{\mu}$
flux models are from pion (solid lines) and kaon (dashed lines) decays
in the SN jet as in Fig. \ref{fig:nuflux}. The two kaon decay models
are due to the protons in the SN jet producing kaons in their whole
(thick dashed lines) and partial (thin dashed lines) energy range (see
Fig. \ref{fig:proton-cool-time}). The lower and upper sets of lines at
$E_{\mu} = 10^3$ GeV correspond to the SN in M82 and NGC253
respectively. The timing uncertainty in the optical detection of SN is
$\sim 1$ day. The corresponding atmospheric background events within
$1^{\circ}$ angular resolution is 0.07 for both M82 and NGC253 in the
same energy range as in Fig. \ref{fig:nuevents}.

Neutrino fluxes of all three flavors ($\nu_{\mu}$, $\nu_{e}$,
$\nu_{\tau}$) on Earth should be equal because of oscillations over
astrophysical distances.  However, only $\nu_{\mu}$'s are emitted from
the sources under consideration and the total number of neutrino
events (including $\nu_{e}$ and $\nu_{\tau}$) will remain the same as
the $\nu_{\mu}$ events calculated here. The lack of good directional
sensitivity for the $\nu_{e}$ and $\nu_{\tau}$ events may prevent
obtaining their positional coincidence with the SN. Timing
coincidences of $\nu_{e}$ and $\nu_{\tau}$ events with $\nu_{\mu}$
events, however, may still be useful to verify the neutrino
oscillations at these energies and test their common origin.

At the quoted rate for M82 and NGC253, a SN from one of these galaxies
would be expected within five years. However, only 1/5 of them would
be {\it visible} in neutrinos due to the $1/2\Gamma_b^2$ beaming
effect of the jet. IceCube may detect one such SN in 25 years of its
operation. The real situation is not as pessimistic as for the
brightest neutrino events expected from M82 and NGC253. Other nearby
spiral galaxies (M31, M74, M51, M101, etc, and the Virgo cluster) will
also have SNe. For a hypothetical SN at 20 Mpc with its jet pointing
towards Earth, the number of neutrino events is $\approx 15$/km$^2$
from pion and kaon decay fluxes combined. With the suggested rate of
$\gtrsim 1$ SN/yr within 20 Mpc, IceCube may detect $\gtrsim 3$ muon
events/yr after beaming correction from the jetted SNe.

\section{Conclusions} \label{sec:summary}

While a core collapse SN in a typical galaxy is a rare event, their
rate of occurence within 20 Mpc could be more than 1/yr, and the
physics and astrophysics one can learn from such an explosion is
enormous. A buried slow jet from the collapsing core of the supernova
progenitor star is an attractive possibility for solving the long
standing problem of how to understand the ejection of the envelope in
SN explosions, by re-energizing the shock wave through energy
deposition by the jet.  Alternatively it could be that only a fraction
of core collapses leads to such jets. These hypothetical mildly
relativistic jets in SNe may be related to the ultra relativistic jets
thought to be responsible for long duration GRBs, which are thought to
originate from the core collapse of massive progenitor stars, some of
which have been positively associated with observed envelope ejection
supernova events.  While all typical core collapses should produce 10
MeV thermal neutrinos, the presence of a jet would allow proton
acceleration by shocks, and produce 1 TeV non-thermal
neutrinos. Detection of these high energy neutrinos by upcoming
Cherenkov detectors would be a smoking gun signal of a SN jet, and
would allow one to study the conditions inside a collapsing star which
may not be possible otherwise.

\section*{Acknowledgments}

Work supported by an NSF grant AST0307376 and NASA grant NAG5-13286.

\end{document}